\documentclass[prl, 10pt, twocolumn, aps, superscriptaddress, floatfix]{revtex4-2}
\usepackage{times, mathrsfs, amsmath, amsfonts, graphics, graphicx, cancel, color, theorem, bbm, mathtools, amssymb}

\usepackage[english]{babel}
\usepackage[margin=.75in]{geometry}
\usepackage[T1]{fontenc}
\usepackage{xfrac,relsize,float}
\usepackage[unicode=true, breaklinks=false, pdfborder={0 0 1}, backref=false, colorlinks=true, linkcolor=blue, citecolor=blue]{hyperref}
\usepackage{physics}

\newcommand{\units}[1]{\,{\rm #1}}

\usepackage{ulem} 
\newcommand{\rem}[1]{}
\newcommand{\com}[1]{}

\begin{document}

\title{Principles underlying efficient exciton transport unveiled by information-geometric analysis}

\author{Scott Davidson}
\email{sd109@hw.ac.uk}
\affiliation{SUPA, Institute of Photonics and Quantum Sciences, Heriot-Watt University, Edinburgh, EH14 4AS, United Kingdom}
\author{Felix A. Pollock}
\email{felix.pollock@monash.edu}
\affiliation{School of Physics and Astronomy, Monash University, Clayton, Victoria 3800, Australia}
\author{Erik Gauger}
\email{e.gauger@hw.ac.uk}
\affiliation{SUPA, Institute of Photonics and Quantum Sciences, Heriot-Watt University, Edinburgh, EH14 4AS, United Kingdom}

\begin{abstract}
\setlength{\parindent}{0em}
Adapting techniques from the field of information geometry, we show that open quantum systems models of Frenkel exciton transport, a prevalent process in photosynthetic networks, belong to a class of mathematical models known as `sloppy'. Performing a Fisher-information-based multi-parameter sensitivity analysis to investigate the full dynamical evolution of the system and reveal this sloppiness, we establish which features of a transport network lie at the heart of efficient performance. We find that fine tuning the excitation energies in the network is generally far more important than optimizing the network geometry and that these conclusions hold for different measures of efficiency and when model parameters are subject to disorder within parameter regimes typical of molecular complexes involved in photosynthesis. Our approach and insights are equally applicable to other physical implementations of quantum transport.
\end{abstract}
\maketitle

\textit{Introduction --}
Energy transport processes are prevalent across science and engineering, from biochemical reactions supporting life~\cite{Intro:PhotosythesisMech, Intro:PhotoExcitons} to nowadays ubiquitous electronic devices~\cite{Intro:Pop, Intro:SolarCells}. Typically, the goal of this process is to guide an excitation from a `source' to a `sink' as quickly and reliably as possible; however, a trade-off between quick versus robust transport often arises. Recently, there has been increasing interest in nanoscale energy transport due to the impressive efficiency of photosynthetic processes~\cite{QT:EfficientPhoto}, potential enhancements in solar cell efficiency~\cite{Intro:CoherentPhotoVoltaics, OQS:Will-GuideSlide}, and for other nanotechnology applications~\cite{Intro:NanotechApplication}. Any practical technology will need to overcome many uncontrollable degrees of freedom, arising from vibrations, the electromagnetic environment, and various quantum mechanical effects. Open quantum systems theory is thus an ideal basis for modelling these transport processes~\cite{OQS:BreuerPetruccione, QT:TransportModels, QT:BuchleitnerReview, Head-MarsdenMazziottiPRA2019}.

Photosynthetic exciton transport networks have inspired much theoretical work investigating the interplay between unitary time evolution and environment-induced decoherence~\cite{QT:OriginalENAQT, QT:MomentumRejuv, QT:UniversalOriginENAQT, QT:QuantumClassicalCrossover, QT:OptimalNoisyQuantumWalks, QT:JianshuDiffusiveTimescales}. Many previous studies have focused on the relationship between network \textit{geometry} (via position-dependent dipole couplings) and transport efficiency: e.g.,~for purely coherent transport, Ref.~\cite{QT:BackbonePair} reports statistical correlations between robust transport and geometries with a `backbone-pair' structure. Similarly, Refs.~\cite{QT:OptimalNetworks, QT:EffCohTransport, QT:Centrosymm} compare the effects of coherent transport, phenomenological site-basis pure dephasing, and network geometry. In the above cases the on-site energies were degenerate and assumed to be unimportant. By contrast,  non-degenerate site energies were used for investigating statistical correlations between dynamics and network design~\cite{QT:MomentInertia, ClaridgeFaraday2020}, and studying optimal trapping processes~\cite{QT:JianshuOptTrapping}. In these studies, pure dephasing serves to overcome (Anderson) localization effects~\cite{QT:CoherenceAndLocalization}. Moving beyond pure dephasing, Ref.~\cite{QT:GeomEffects}, investigates geometrical effects with a more realistic environment model, whereas Refs.~\cite{OQS:ChinVibra, OQS:IrishVibra} incorporate vibronic effects. Finally, Ref.~\cite{QT:BennettVibraRobustness} analyzes transport efficiency and robustness in different vibrational parameter regimes with fixed, non-degenerate on-site energies.

\begin{figure}[t]
	\includegraphics[scale=0.31]{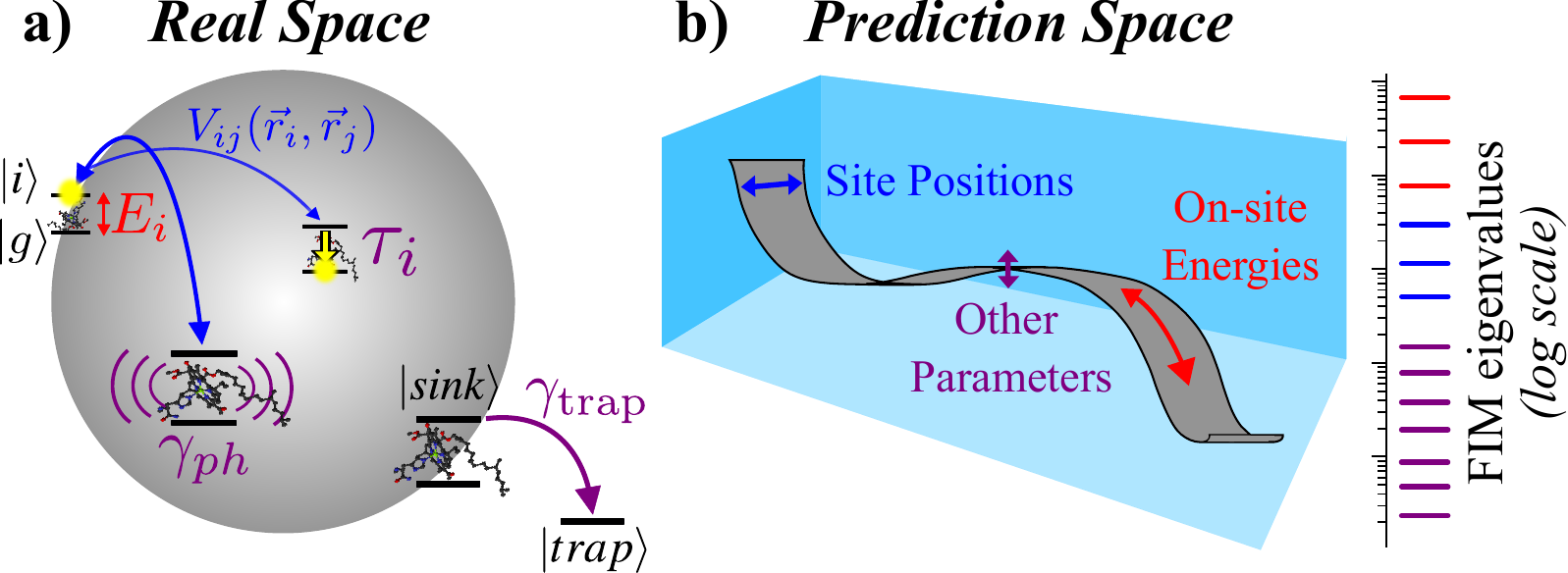}
    \caption{
	\textbf{a)}  Schematic depiction of the nanoscale transport systems considered  here. Our models account for coherent transport ($V_{ij}$), vibrational coupling ($\gamma_{ph}$), spontaneous decay ($\tau_i$) and energy extraction ($\gamma_\text{trap}$). \textbf{b)}  Cartoon depiction of the parameter space manifold revealed by our information-geometric analysis and the associated log-linear (`sloppy') Fisher Information Matrix (FIM) eigenvalue spectrum which suggests that on-site excitation energies are the dominant parameters in determining transport performance.
	}
	\label{fig:IntroFig}
\end{figure}

In this Letter,  we use information-geometric tools to show that, with realistic noise models, on-site energy tuning is typically far more important to transport performance than spatial configuration. While we focus on (Frenkel) exciton transport through a small network of  sites, in a regime appropriate to the early stages of photosynthesis, our results are  equally relevant to the design of organic photovoltaics, molecular charge transfer systems, and quantum walks more generally~\cite{QT:JakubSingleMolecule, QT:OrganicSolarCells, QT:QuantumWalksReview}. Constructing the Fisher Information Matrix (FIM) for our transport model, we show that its parameter space possesses a surprising structure which is common to mathematical models across  scientific disciplines from biology to machine learning. These models have been termed `sloppy' since they meaningfully depend on only a few key parameter combinations~\cite{Sl:SloppinessReview}. We proceed by exploiting this property to analyse which \textit{physical} parameters must be controlled most carefully in order to realize efficient quantum transport. Finally, we investigate how variations in both the transport model and employed efficiency measure affect this parameter sensitivity.  Going beyond phenomenologically informed previous works, we elucidate how details of the vibrational environment determine whether the energetic landscape or network geometry most significantly influence the system's transport properties, confirming that in molecular networks the former typically dominates.

\textit{Transport model --}
We consider a network comprising $N$ sites, each modelled as a two-level system, and assume there is never more than a single excitation present. The unitary dynamics are governed by the Hamiltonian
\begin{equation}
	\hat{\mathcal{H}}_{s} = \sum_{i=1}^N E_i \ket{i}\!\bra{i} + \sum_{i \neq j} V_{ij}   \left( \ket{i}\!\bra{j} + \ket{j}\!\bra{i} \right)~,
	\label{eq:Unitary}
\end{equation}
with inter-site coupling strength $V_{ij}$ between site-basis states $\ket{i}$ and $\ket{j}$, and  on-site excitation energy $E_i$ relative to ground state $\ket{g}$. For simplicity, we present the case $N=4$; however, as shown in the Supplemental Material (SM)~\footnote{See Supplemental Material for this Letter at }, our conclusions hold for other small networks. The dipolar hopping terms in Eq.~\eqref{eq:Unitary} are given by $V_{ij}(\vec{r}_i, \vec{r}_j) = D / |\vec{r}_i - \vec{r}_j|^3$, where $\vec{r}_i$ denotes the real-space coordinates of site $\ket{i}$ and $D$ is a coupling constant,   assumed equal for all sites.

Additionally, we account for a finite temperature phonon environment, which couples to our transport network through site-dependent displacements with interaction Hamiltonian $\hat{\mathcal{H}}_{I,\ phonon} = \sum_i^N \ket{i}\!\bra{i} \otimes \sum_k g_k ( \hat{c}_{i,k} + \hat{c}^\dagger_{i,k} )$, where $\hat{c}^{(\dagger)}_{i,k}$ is the annihilation (creation) operator of local phonon mode $k$ and $g_k$ is the associated coupling constant. For computational convenience, we assume weak to moderate system-environment couplings that are adequately described with a  Bloch-Redfield master equation~\cite{OQS:BreuerPetruccione}. This vibrational environment  leads to phonon-mediated eigenstate transitions with rates determined by the phonon spectral density $J(\omega)$~\cite{OQS:GaugerHeatPumping}. 

We focus on a structured, singly-peaked spectrum of the form
\begin{equation}
    J_1(\omega) = \gamma_{ph} \left|\frac{\omega}{\omega_c}\right|^3 e^{-\left( \frac{\omega}{\omega_c}\right)^2} \left[ n_{ \rm BE}(\omega, T_{ph}) + \Theta(\omega) \right] \label{eq:Jw1},
\end{equation}
where $\gamma_{ph}$ controls the overall system-phonon coupling, the Bose-Einstein distribution $n_{ \rm BE}$ describes a phonon environment at temperature $T_{ph}$, the Heaviside step function $\Theta(\omega)$ reflects the prevalence of phonon emission over absorption at finite temperature and $\omega_c$ is a high frequency cutoff. This generic form captures the typical situation of a vanishing $J(\omega)$ in both  low and high frequency limits. Additionally, we consider two other functional forms; namely $J_2(\omega) = \gamma_{ph}\Theta(\omega)$ and $J_3(\omega) = \gamma_{ph}$, to investigate which features of $J_1$ are critical to our findings. Physically, $J_3$ represents a flat, infinite temperature spectrum which, while unrealistic in practice, leads to Markovian pure dephasing (used in many previous exciton transport studies~\cite{QT:OptimalNetworks, QT:EffCohTransport, QT:Centrosymm, QT:MomentInertia}) within our non-secular Bloch-Redfield treatment; thus allowing for comparison with existing literature. Spectrum $J_2$ describes a flat, \textit{zero} temperature environment and acts as bridge between $J_1$ and $J_3$ by retaining the thermodynamically consistent asymmetry between phonon absorption and emission while removing all other structure.

Finally, we account for two other physical processes of relevance to energy transport phenomena. After successful transport, the Lindblad operator $\hat{{L}}_{t} = {\gamma_\text{trap}}^{1/2} \ket{\text{trap}}\!\bra{\text{sink}}$  incoherently extracts an excitation from the designated network sink to an orthogonal trap state (Fig.~\ref{fig:IntroFig}a). We also include spontaneous  decay via $\hat{{L}}_{d,i} = {\tau_i}^{-1/2} \ket{g}\!\bra{i}$, which destroys an excitation at site $\ket{i}$.

Overall, the transport dynamics of our model is governed by the Bloch-Redfield equation:
\begin{equation}
    \frac{\partial\rho_s}{\partial t} = -i[\hat{\mathcal{H}}_{s}, \rho_s] + \hat{\mathcal{R}}\rho_s + \hat{\mathcal{L}}[\hat{L}_t]\rho_s + \sum_{i=1}^N \hat{\mathcal{L}}[\hat{L}_{d,i}] \rho_s ~,
    \label{eq:brme}
\end{equation}
where subscript $s$ denotes system quantities, $\hat{\mathcal{R}}$ is our phonon interaction Bloch-Redfield tensor, and $\hat{\mathcal{L}}[\bullet]\rho$ is the usual Lindblad dissipator~\cite{OQS:BreuerPetruccione}. 

As a means to parameterize our transport networks in terms of \textit{physically} relevant quantities (see Fig.~\ref{fig:IntroFig}), we use the following set of model parameters throughout this work: on-site excitation energies $E_{i}$; exciton lifetimes $\tau_i$; real-space polar co-ordinates $\vec{r}_{1i} = (r_{1i}, \phi_{1i}, \theta_{1i})$ w.r.t.~site $\ket{1}$; system to trap extraction rate $\gamma_\text{trap}$; system-environment vibrational coupling $\gamma_{ph}$, and phonon bath temperature $T_{ph}$. We denote the vector containing all of these input parameters as $\vec{\eta} = \{\eta_\mu\}$. Additional parameterization details and parameter values for our results are provided in the SM~\cite{Note1}.

Given sufficient computational resources, the ensuing analysis could instead utilize other, more accurate numerical methods for simulating open quantum dynamics~\cite{OQS:DeVegaReview} such as polaron-transform based master equations~\cite{OQS:VarME, OQS:OlayaCastroPolaronME, OQS:JangPolaronME}, Hierarchical Equations of Motion~\cite{OQS:HEOM, KaisScaledHEOM}, time-dependent Density Matrix Renormalization Group approaches~\cite{OQS:Plenio-td-DMRG,OQS:SomozaDAMPF} or path integral based approaches~\cite{OQS:StrathearnNMPIDynamics, OQS:TEMPO, OQS:PollockCausalTensorNetworkPI}. More broadly, 
our approach is equally applicable to far more complex models of physical systems than the simple tight-binding models considered here.

\begin{figure}
	\includegraphics[scale=0.49]{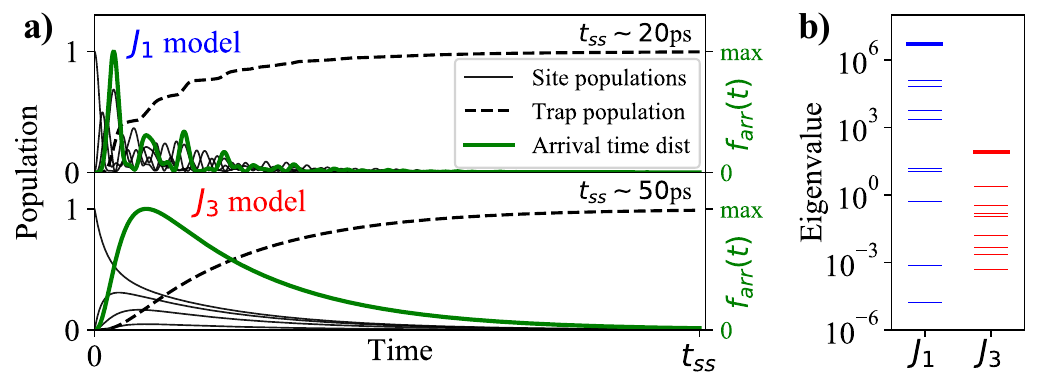}
	\caption{
	Comparison between transport models with $J_1$ (phonon-mediated eigenstates transitions) and $J_3$ (site-basis pure dephasing) vibrational environments with all other parameters identical. \textbf{(a)} Population dynamics up to respective steady state times ($t_{ss}$) for each system site (black lines) and trap state (dashed line) with associated arrival time distribution ($f_\text{arr}$ -- green) on right-hand axis; \textbf{(b)} sloppiness of FIM eigenvalue spectra.
	}
	\label{fig:SloppyDynamicsFig}
\end{figure}

\textit{Fisher Information Matrix --}
 Our primary focus is the non-equilibrium transport properties of a single energy quantum that is generated, transported and extracted before another excitation enters the system; reflecting excitation events that are rare or transient. However, as shown later, our general conclusions also hold in steady-state scenarios. For now, to investigate transport efficiency over the full time evolution of the system, we consider the probability distribution $f_{arr}(t, \vec{\eta}) = (P_{max})^{-1} \frac{\partial}{\partial t} \bra{\text{trap}} \rho(t, \vec{\eta}) \ket{\text{trap}}$, where $P_{max}$ is the final trap population. This `arrival time' distribution is the likelihood of the excitation arriving at the trap at time $t$, normalized by the total probability of successful transport. Using this, measures of transport efficiency such as the mean and variance of the arrival time can be computed. Our goal is to determine which, and how many, of our model's physical input parameters ($\vec{\eta}$) contribute most to determining these properties. We explore the parameter space of our models with the Fisher Information Matrix (FIM)~\cite{Sl:InfoGeomMethods} which, for a distribution $f(t, \vec{\eta})$, is defined as
\begin{equation}
	g_{\mu\nu}(\vec{\eta}) = E\bigg[ \bigg( \frac{\partial}{\partial \eta_\mu} \ln f(t, \vec{\eta} ) \bigg) \bigg( \frac{\partial}{\partial \eta_\nu} \ln f(t, \vec{\eta} ) \bigg) \bigg|\ t\ \bigg] ~,
	\label{eq:FIM}
\end{equation}
where $E[\ ...\ |\ t\ ]$ denotes the expectation value w.r.t.~time. The FIM \eqref{eq:FIM} is a metric tensor  for the parameter space manifold (i.e. subspace of possible model outputs) embedded within the full `prediction space' of possible distributions~\cite{Sl:SloppinessReview}. Following \cite{Sl:DufresneGeomOfSloppiness}, an intuitive geometric structure emerges from the FIM,  in which the manifold possesses a hierarchy of widths (i.e.~geodesic lengths) corresponding to successive, \textit{exponentially} less important linear combinations of model  parameters (see Fig.~\ref{fig:IntroFig}). Similar `sloppy' geometric structures are present in many unrelated mathematical models, from a variety of scientific disciplines~\cite{Sl:ScienceSloppiness, Sl:SloppyUniversalityClass, Sl:OriginalBioSloppiness}. Formally, the sloppiness of a particular model is determined by the manifold's  widths along each of the FIM eigenvector directions. However, following Transtrum \textit{et.~al.}~\cite{Sl:SloppinessGeometry, Sl:SloppinessReview}, the FIM eigenvalues may be used as the key indicator of model sloppiness. Throughout this work we construct the FIM using dimensionless (\textit{log}) parameters to allow for fair comparison between parameters with different physical dimensions (see SM~\cite{Note1} for details).

\textit{Results \& Discussion --}
We begin by investigating how different vibrational environment descriptions affect the robustness of transport properties, as quantified by the sloppiness of the model. Specifically, we compare spectra $J_1$ and $J_3$ for a transport model with linear chain geometry and degenerate on-site energies, finding  that this  leads to two contrasting  parameter space manifolds, where the transport properties are several orders of magnitude more sensitive to parameter perturbations for the first model ($J_1$) than the second~($J_3$); as seen by the relative difference in FIM eigenvalues. In the SM~\cite{Note1} we show that this contrast holds over a wide range of $\gamma_{ph}$ values.  The population dynamics in Fig.~\ref{fig:SloppyDynamicsFig}a show that only the $J_1$  model supports coherent (site-basis) oscillations on timescales  relevant to the transport process. In this case, the resulting interference effects contribute to a dynamics that is far more sensitive to the details of the model. However, both models are inherently sloppy; sensitivity to small parameter changes is concentrated in a few linearly independent perturbations, which we will now see primarily involve the on-site excitation energies.

We proceed by analysing the FIM eigenvectors to perform a sensitivity analysis across different parameter regimes. This tells us which physical model parameters have the largest influence on the dynamical evolution of the system. The linear combination of bare model parameters which forms each eigenvector, paired with the magnitude of its eigenvalue, establishes the {\it local} (in parameter space) hierarchy of parameter sensitivities.  In the SM~\cite{Note1} we show that, due to the inherent complexity of these transport networks, and the competing influences of  various parameters, the parameter space lacks clear global patterns in these important linear combinations. Therefore, we instead look for bare parameters which consistently dominate the eigenvector(s) with the largest  eigenvalue(s) (i.e.~we identify a lower-dimensional parameter subspace which influences transport behaviour significantly). To quantify relative parameter importance we use $\mathcal{P}({\eta_\mu}) = \sum_i \lambda_i\ |\hat{e}_i \cdot \hat{\eta}_\mu|$ and normalize such that $\sum_\mu \mathcal{P}(\eta_\mu) = 1$, where  $\hat{e}_i$ ($\lambda_i$) is the $i^{th}$ eigenvector (eigenvalue) of the FIM, and $\hat{\eta}_\mu$ is the parameter space basis vector for model parameter $\eta_\mu$. We use absolute values since we are concerned with the magnitude  of the sensitivity. This method establishes the relative parameter importance over the full dynamical transport process instead of focusing on sensitivity of a single scalar quantity.

\begin{figure}
	\includegraphics[scale=0.39]{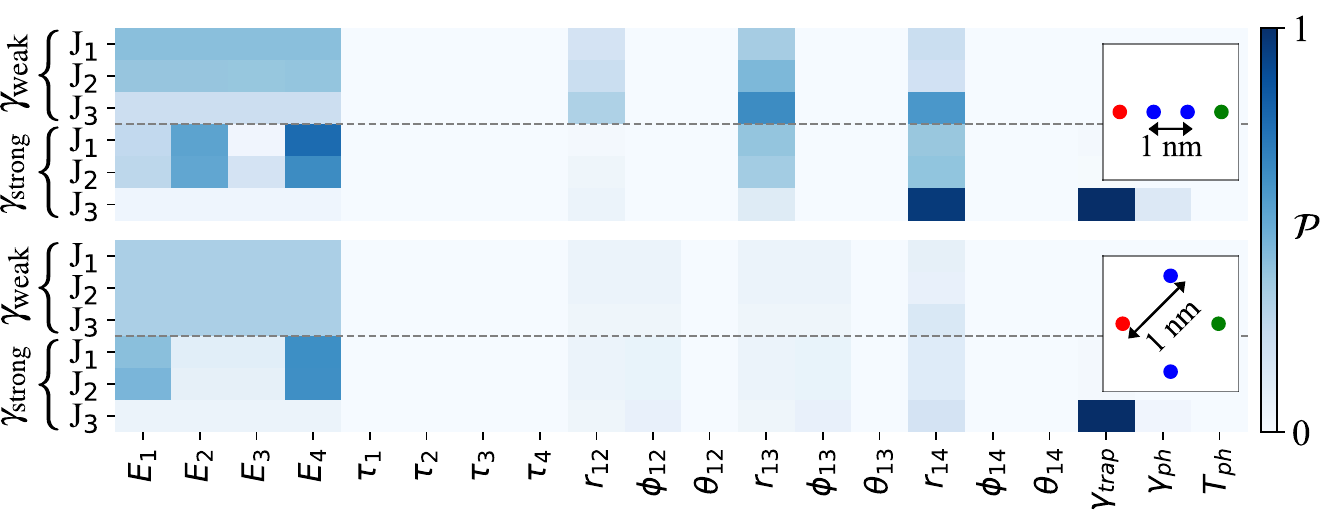}
	\caption{
	Relative importance $\mathcal{P}(\eta_\mu)$ of each model parameter $\eta_\mu$ for  simple 1D and 2D transport geometries. Rows correspond to different functional forms of phonon spectrum (Eq.~\ref{eq:Jw1}) for both weak ($\gamma_\text{weak} = 10^{-3}\units{eV}$) and stronger ($\gamma_\text{strong} = 10^{-1}\units{eV}$) phonon coupling ($\gamma_{ph}$). Insets depict network geometries with red (green) dots representing the source (sink) site. 
	}
	\label{fig:GeometryVariation}
\end{figure}

The quantity $\mathcal{P}(\eta_\mu)$ is shown in Fig.~\ref{fig:GeometryVariation} for each physical input parameter ($\vec{\eta}$) in simple 1D and 2D geometries with  different phonon environments.  For weak system-phonon coupling, the parameters which control the system's unitary dynamics (i.e.~positions and on-site energies) are significantly more important. Furthermore, we see a close similarity between $J_1$ and $J_2$ but,  for pure-dephasing-like noise ($J_3$), there are \textit{qualitative} differences in parameter importance. This suggests that the presence of an asymmetry in phonon absorption and emission rates, derived from the thermodynamic consistency of models with spectra $J_1$ and $J_2$, is a key feature in determining which parameters are most important in exciton transport. Physically, a phonon rate asymmetry leads to lower energy eigenstates becoming increasingly populated as the evolution progresses. Therefore, the precise properties of these eigenstates will become more important. Without this asymmetry, and in the absence of an extraction process, the system would tend towards a maximally mixed state with equal population on all sites. In that case, the precise details of the system Hamiltonian and its resulting eigenstates will be less important, and site positions and excitation energies will have less influence on the system's dynamics. Finally, we note that stronger phonon coupling tends to exaggerate the above-discussed effects: for $J_1, J_2$, the lower energy states are populated more quickly and so the details of these states, and the different on-site energies and positions which determine them, are more important. Similarly, for strong pure dephasing, the maximally mixed state is reached more quickly; therefore, the most important considerations are simply the extraction rate ($\gamma_\text{trap}$) and the proximity of the sink site to all others.

\begin{figure}
	\includegraphics[scale=0.31]{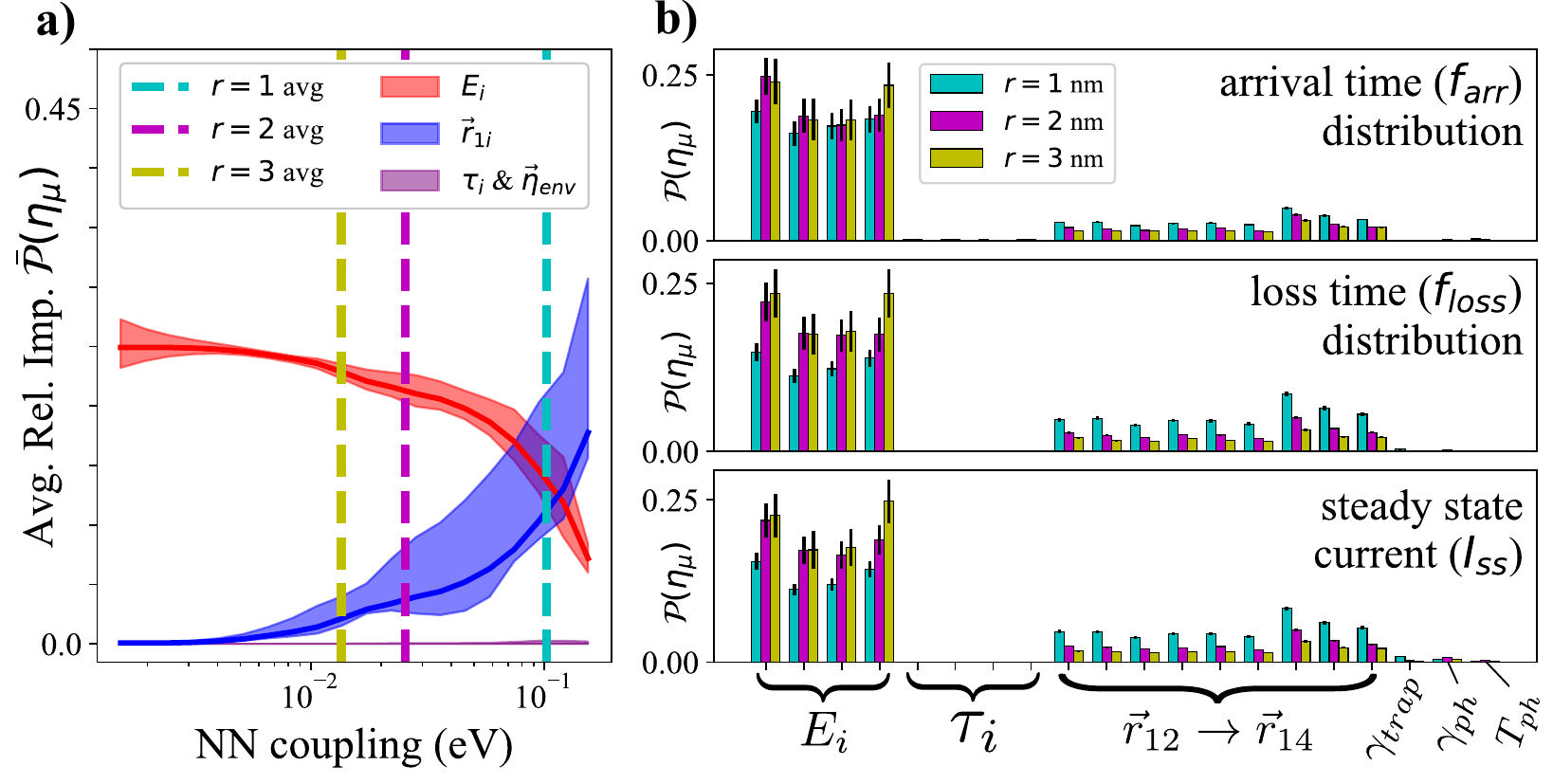}
	\caption{
	{\bf (a)} Average (solid lines) and range (shaded area) of relative importance for different parameter \textit{groups} upon varying site spacing in a linear chain geometry (i.e.~all pairwise couplings change but  nearest neighbour couplings shown for simplicity). {\bf (b)} Average parameter importance (main bar) and associated variance (error bar) for different transport efficiency measures using randomly generated four-site networks (1000 per $r$ value). Geometries were constrained within a sphere of radius $r$ with 10\% disorder applied to all other model parameters (see main text and SM~\cite{Note1}).  Dashed vertical lines in (a) denote average nearest neighbour coupling   across all networks for the each of the three different sphere radii in (b).
	}
	\label{fig:Robustness}
\end{figure}

For the remainder of this Letter, we focus on the  $J_1$ phonon environment and turn to look more closely at the Hamiltonian parameters. We find that there is a connection between coherent coupling strength and the relative importance of network geometry vs.~on-site energies. Figure~\ref{fig:Robustness}a compares these two parameter groups for a degenerate linear chain, showing that network geometry becomes more important as  inter-site coupling increases. However, on-site energies  are dominant up to nearest neighbour hopping around $100\units{meV}$. This can be explained as follows: as coherent couplings increase, eigenenergy splittings also increase. For structured spectral densities, such as $J_1$, this entails larger differences between rates of the various phonon-mediated eigenstate transitions and therefore different transport performance. In the SM~\cite{Note1} we show that this energy-position importance crossover disappears for pure-dephasing models. The shaded regions in Fig.~\ref{fig:Robustness}a are bounded by the  maximal/minimal values of $\mathcal{P(\eta_\mu)}$ when restricting $\vec{\eta}$ to encompass only the corresponding parameter group.

Figure~\ref{fig:Robustness}b corroborates the dominant importance of on-site energies across a large region of parameter space. Its panels show the \textit{average} relative parameter importance for randomly generated network geometries inside a sphere of radius $r$ with source and sink sites fixed at opposite poles and intermediate sites randomly positioned (see Fig.~\ref{fig:IntroFig}a), as in Refs.~\cite{QT:Centrosymm, QT:OptimalNetworks, QT:EffCohTransport}. Consequently, the average coherent coupling strength is inversely related to $r$. Random disorder is also applied to all other parameters (see SM~\cite{Note1} for procedural details). In all cases,  we focus on networks exhibiting relatively efficient transport by restricting to randomly generated networks whose time evolution lasts less than $1\units{ns}$. The top panel shows results for the arrival time distribution $f_{arr}(t, \vec{\eta})$. The middle panel uses an alternative indicator of successful transport, namely the `{\it loss time}' distribution $f_{loss}(t, \vec{\eta}) = (P_{loss})^{-1} \frac{\partial}{\partial t} \bra{g} \rho(t, \vec{\eta}) \ket{g}$ which describes the probability of exciton recombination at time $t$, normalized by the overall recombination probability $P_{loss}$. Finally, with previous studies arguing that, for photosynthetic systems under incoherent illumination, only steady-state transport properties are relevant~\cite{QT:KassalCoherentClassification,QT:UniversalOriginENAQT}, we also consider the steady-state current into the trap, $I_{ss} = \gamma_\text{trap} \bra{{\rm sink}} \rho_{ss}(\vec{\eta}) \ket{{\rm sink}}$  which necessitates adding an injection term $\mathcal{\hat{L}}[\sqrt{\gamma_{inj}}\hat{\sigma}^{+}_1] \rho_s$ to Eq.~\eqref{eq:brme} for re-populating the source site. Here, we simply look at the magnitudes of the partial derivatives ($|\partial_\mu I_{ss}|$) for parameter sensitivities. However, our non-trivial information-geometric approach could  be adapted to steady state scenarios by choosing $f(\vec{\eta})$ in Eq.~\eqref{eq:FIM} to be the full counting distribution of transported excitons~\cite{QT:EmaryCountingStat}. 

As shown in Fig.~\ref{fig:Robustness}b, upon moving away from the idealized degenerate linear chain towards disordered model parameters, the importance of tuning the on-site energies dominates that of network geometry, regardless of the adopted efficiency measure. By comparing with the coloured vertical lines in Fig.~\ref{fig:Robustness}a, we see that the transition from importance of on-site energies to network geometries at stronger couplings -- as seen in panel (a) -- no longer occurs, even for $r=1\units{nm}$, where average nearest neighbour couplings approach $100\units{meV}$. Whilst all three efficiency measures show positions becoming slightly more important with decreasing sphere radii, we nonetheless conclude that, on average, tuning the on-site energies remains far more important for achieving efficient transport . 

Finally, since the magnitude of inter-site couplings in typical photosynthetic system is $\lesssim 10 \units{meV}$, our results suggest that on-site energy tuning, rather than molecular geometry, is primarily responsible for the impressive efficiency of natural photosynthetic processes. See SM~\cite{Note1} for an extended discussion of experimental platforms to which our results apply.

\textit{Conclusion --}
We have demonstrated that open quantum systems models of exciton transport are `\textit{sloppy}', in common with many other multi-parameter models from various scientific domains. For the generic transport models studied here, we found the extent of the sloppiness to depend on the nature of the phonon environment and, in particular, the coupling strength and finite temperature asymmetry in phonon rates. Making use of the intrinsic hierarchy of parameter sensitivities implied by the sloppiness of our model, we performed an information-geometric multi-parameter sensitivity analysis to understand which physical features of the system most affect its transport performance. We anticipate that the methodology introduced here could become an important asset for analysing a host of other settings involving complex quantum dynamics, complementing other approaches that aim to capture its key features~\cite{arXiv:2009.03902, GuoModiPRA2020, LuchnikovPRL2020}. Crucially, it also opens the door to the application of other information-geometric analyses~\cite{Sl:MBAM, Sl:SloppinessGeometry, Sl:SloppyUniversalityClass}.

Our results underline the importance of a tailored on-site energy landscape for achieving efficient quantum transport \cite{ClaridgeFaraday2020} and demonstrate considerable robustness to parameter disorder. Intuitively, this parameter importance reflects the fact that on-site energies have a direct influence on both the unitary and non-unitary dynamics through their effects on the eigenenergies and eigenstates of the system, therefore governing coherent as well as dissipative processes. Our findings suggest the future exploration of non-trivial energy landscapes, such as those recently found in~\cite{SD-JCP-biased-chains}, could lead to significant improvements in our understanding and implementation of efficient quantum transport processes.

\begin{acknowledgments}
{\it Acknowledgements --} SD was supported by the EPSRC Grant No.~EP/L015110/1. EMG acknowledges support from the Royal Society of Edinburgh and Scottish Government and EPSRC Grant No. EP/T007214/1. Computations were performed using QuantumOptics.jl~\cite{QO.jl}.
\end{acknowledgments}

%
%

\hfill \linebreak \hfill \linebreak \hfill \linebreak

\section*{\textbf{Supplemental Material}}

\renewcommand{\thefigure}{S\arabic{figure}} 

\appendix

\section{Model Parametrization}

In this section, we provide some details on our transport model parametrization and how these parameters are used in the FIM. Each of the sites in our transport network is parameterized by an on-site energy~$E_i$, an intrinsic decay lifetime~$\tau_i$, and a real space site position. The positions are specified in spherical coordinates with the `source' site fixed at the origin so that sites $2$ to $N$ have coordinates $(r_{1i}, \theta_{1i}, \phi_{1i})$, which denote their position relative to the source. In order to characterize the vibrational system-environmental coupling in our model, we use an overall system-phonon coupling strength $\gamma_{ph}$ and an environment temperature $T_{ph}$. To describe the exciton extraction process we use a rate $\gamma_{trap}$. This means that all our environmental processes are captured in these three variables. Combining all of these, for a four-site network, we have a vector of parameters given by
\begin{align}
    \vec{\eta} = \bigg\{&\ E_1,\ E_2,\ E_3,\ E_4,\ \tau_1,\ \tau_2,\ \tau_3,\ \tau_4, \notag \\
    &r_{12},\ \theta_{12},\ \phi_{12},\ r_{13},\ \theta_{13},\ \phi_{13},\ r_{14},\ \theta_{14},\  \phi_{14},  \notag \\ 
    &\gamma_{trap},\ \gamma_{ph},\ T_{ph} \bigg\}, 
    \label{eq:eta-vec}
\end{align}
with respect to which we then take derivatives in order to form the FIM. For example, the matrix element $g_{12}$ is given by
\begin{equation}
    g_{12} = \int_0^\infty dt \frac{1}{f(t, \vec{\eta})} \frac{\partial f}{\partial E_1} \frac{\partial f}{\partial E_2}, 
\end{equation}
where $f$ is the either the `\textit{arrival-time}' or `\textit{loss-time}' 
probability density function defined in the main text. All of the partial derivatives and the subsequent integration are performed using numerical methods. Due to the factor of $1/f$ in the integrand, the computations can be susceptible to numerical noise in some cases. To overcome this in our implementation, we cut off any parts of the integrand which are dominated by numerical noise.

As mentioned in the main text, we use a $\log$-parametrization in order to make the FIM dimensionless. Explicitly, if we let $\tilde{\eta} = \log \eta$, then the differential element describes a \textit{relative} change in the parameter
\begin{equation}
    d\tilde{\eta} = \frac{d\eta}{\eta},
\end{equation}
rather than an absolute change. Furthermore, for an arbitrary probability distribution $y(\eta)$, we can write
\begin{equation}
    \frac{\partial y}{\partial \tilde{\eta}} = \eta \frac{\partial y}{\partial \eta},
    \label{eq:logDeriv}
\end{equation}
which is a dimensionless quantity. We use this procedure for all model parameters except the angular co-ordinates of each site, since these angles are already dimensionless quantities and can have an arbitrary factor of $\{2n\pi : n \in \mathbbm{Z} \}$ added to them; this would have the effect of artificially `inflating' the logarithmic angular derivatives due to the factor of $\eta$ in Eq.~\eqref{eq:logDeriv}.

\section{Default Transport Parameters}

Table~\ref{tab:param-vals} lists the default set of transport model parameters used throughout the main text and SM unless otherwise specified. The value of the dipole coupling constant $D$ is arbitrarily chosen such that dipoles separated by $1$ or $2\units{nm}$ lead to couplings in the $10$'s of meV; as is found in typical organic systems (see `Applicable Physical Systems' section below). Our choice of typical exciton lifetime $\tau_i$ is also phenomenologically motivated. In practice, the decay of an exciton could feasibly occur via radiative or non-radiative recombination. Rates for these processes can vary significantly~\cite{Intro:PhotosythesisMech,QT:BrickerDecay}, and here we simply use an overall exciton lifetime $\tau_i = 10\units{ns}$ at each site (at the long end of the plausible spectrum) to characterize recombination processes. This provides an alternate loss channel which penalizes particularly slow transport. In main text Fig.~2 we use a linear chain geometry with NN spacing at $3\units{nm}$ for both $J_1$ and $J_3$ models.

\begin{table}
    \begin{tabular}{ |c|c|c|c| }
        \hline
        \textbf{Parameter} & \textbf{Value} & \textbf{Parameter} & \textbf{Value} \\ \hline
        $E_{i}$ & $2\units{eV}$ & $D$ & $10^{-8} \units{eV^{-2}}$ \\ 
        $\tau_{i}$ & $10\units{ns}$ & *$\gamma_{inj}$ & $10^{-4}\units{eV}$ \\ 
        $T_{ph}$ & $300\units{K}$ & $\gamma_{ph}$ & $10^{-2}\units{eV}$ \\ 
        $\gamma_{trap}$ & $10^{-3}\units{eV}$ & $\omega_c$ & $10^{-1} \units{eV}$\\ 
        \hline
    \end{tabular}
    \caption{Default parameter values used throughout main text. *Injection rate $\gamma_{inj}$ is only non-zero when steady state currents are considered.}
    \label{tab:param-vals}
\end{table}

\section{Applicable Physical Systems}
\label{sec:exp-discuss}

In the main text it was found that the transport properties in a idealized linear chain system were far more sensitive to perturbations in the on-site energies in comparison to the site positions and all other model parameters. This on-site energy dominance persisted up to coherent nearest-neighbour couplings of order $\sim 100 \units{meV}$ for the idealized case, and persisted to even higher coupling strengths when random disorder in all model parameters was included. In this section, we list a number of experimental platforms and discuss the varying extent to which our generic model may be applicable in each of these physical systems.\\

{\setlength{\parindent}{0em} \textit{Natural photosynthetic complexes:}} \\
Perhaps the most relevant application for our model system, these complexes generally consist of a collection of chromophores (`sites' in our model) which engage in exciton generation and transport processes; the goal of which is to funnel excitons across the various chromophores and into a `Reaction Center' (RC) complex where the charge separation can occur. There are a number of well studied photosynthetic systems including the Fenna-Matthews-Olsen complex of \textit{Prosthecochloris aestuarii} and \textit{Chlorobium tepidum} bacteria, as well as the Light Harvesting 1 \& 2 and RC complexes of purple bacteria. Numerical values for the on-site energies and inter-site coherent coupling strengths in each of these examples have been obtained from various experimental observations and fitting procedures~\cite{QT:typical-photosynth-params-1, QT:typical-photosynth-params-2, QT:typical-photosynth-params-3, QT:typical-photosynth-params-4, QT:typical-photosynth-params-5, QT:typical-photosynth-params-6, QT:EfficiencyFromENAQT}; all of which agree that the typical magnitude of coherent couplings in such systems is $\lesssim 10 \units{meV}$. Therefore, our results suggest that it may be the intricate details of the on-site energy disorder in such systems, rather than the real space geometry as previously assumed, which contributes most significantly to the impressive quantum yield of the photosynthesis. \\

{\setlength{\parindent}{0em} \textit{Artificial light-harvesting devices:}} \\
Artificial devices, such as organic photo-voltaics (OPVs), consist of a collection of donor and acceptor molecules. The absorption of a solar photon generates an exciton which can then diffuse to a donor-acceptor interface where charge separation can occur. Since these systems are comprised of a large number of organic molecules, the typical energy scales involved are approximately similar in magnitude to natural photosynthetic complexes. Specifically, nearest neighbour couplings can be of order $50 \sim 100 \units{meV}$~\cite{QT:OPV-hopping-rates-1, QT:OPV-hopping-rates-2, QT:OPV-hopping-rates-3} and significant on-site energy disorder can be present. Furthermore, transport processes in such systems occur over a wide range of length and time scales (from tenths to hundreds of nanometers and from femto to nano seconds respectively), and both transient dynamics and steady state diffusion are thought to be important~\cite{QT:OPV-review, QT:OPV-modelling-review}. Our results are therefore generally applicable in these regimes. However, in systems with large donor/acceptor molecules, strong local and non-local phonon interactions may be relevant~\cite{QT:OPV-modelling-review}. This would likely lead to deviations from our simple Bloch-Redfield model and, as mentioned in the main text, the FIM sensitivity analysis would have to be built atop an alternative method for quantum dynamics simulation to account for stronger system-environment couplings. \\

{\setlength{\parindent}{0em} \textit{Trapped ion chains:}} \\
An entirely different system in which non-trivial quantum transport dynamics have been demonstrated is within linear chains of trapped ions~\cite{QT:ion-chain-ENAQT}. This experimental implementation can be accurately mapped to the single-excitation subspace models used throughout our work and both Markovian [our $J_3(\omega)$ spectrum in the main text] and non-Markovian [e.g.~our $J_1(\omega)$ spectrum] dephasing processes can be engineered in these chains. However, the length and time scales involved differ from our work by many orders of magnitude; therefore, further work is needed to ascertain the extent to which our specific conclusions are directly applicable to this `simulation platform', however, we see no reason why our general approach cannot be applied. \\

{\setlength{\parindent}{0em} \textit{Quantum Dots:}} \\
Finally, quantum dot (QD) systems with sufficiently weak inter-dot coupling (relative to their binding energies) such as CdSe QDs can support exciton states and, in colloidal and superlattice arrangements, can exhibit dynamical, as well as steady state, exciton diffusion/transport processes~\cite{QT:CdSe-QD-superlattices, QT:Colloidal-CdSe-transient-diffusion}. Furthermore, charge transport has been studied in electrostatically defined QDs~\cite{QT:Graphene-QD-hopping-tuning, QT:SiGe-QD-hopping-tuning}. In this case control over energy levels and inter-dot coupling (i.e. `network geometries' in our study) has been demonstrated in a number of QD systems~\cite{QT:Graphene-QD-hopping-tuning, QT:SiGe-QD-hopping-tuning, QT:GaAs-QD-hopping-tuning}, suggesting that our results may also be generally applicable here.

\section{Extended Fig. 2}

\begin{figure}
	\includegraphics[scale=0.42]{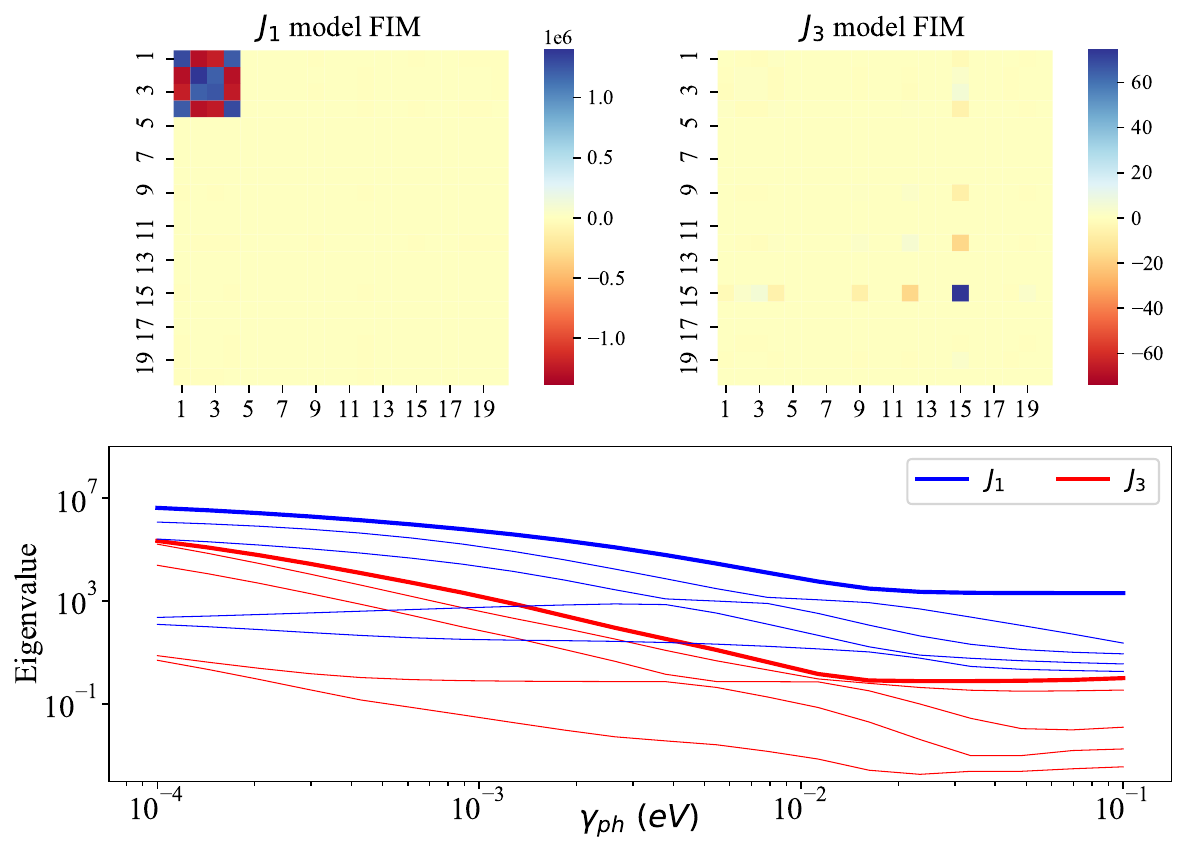}
	\caption{
	\textit{Top} - Fisher Information Matrix (FIM) for the two models discussed in Fig.~2 of the main text. Order of rows/columns correspond to the same order as the parameters are listed in Eq.~\ref{eq:eta-vec} (e.g. row/column 1 = $E_1$, row/column 2 = $E_2$ $\hdots$ row/column 20 = $T_{ph}$). \textit{Bottom} - Five largest FIM eigenvalues for both models over a range of system-phonon coupling strengths.
	}
	\label{fig:ext-fig2}
\end{figure}

Figure~\ref{fig:ext-fig2} is an extension of the main text Fig.~2. It explicitly shows the Fisher Information matrices for the $J_1$ and $J_3$ models discussed in the main text as well as the robustness of the contrast in FIM eigenvalue spectra for the two models across a large range of system-phonon coupling values. The dominance of both the on-site energy related matrix elements in the $J_1$ model FIM, and of the $r_{14}$ related elements in the $J_3$ model FIM are in close agreement with the insights gained from Fig.~3 in the main text. The relative magnitude of the largest FIM eigenvalues for the $J_1$ model compared with the $J_3$ model across a wide range of coupling strengths, as seen in the bottom panel of Fig.~\ref{fig:ext-fig2}, highlights the intrinsic increase in sensitivity to parameter perturbation that arises when a more realistic phonon environment description ($J_1$) is used.

\section{Randomization Procedure}

Here, we briefly describe the process used to generate the random networks for Fig.~4 of the main text (and Fig.~\ref{fig:SI_RobustnessN6}). In order to generate the random geometries in a sphere of radius $r$, we first fix the source and sink sites at Cartesian co-ordinates $(-r, 0, 0)$ and $(r, 0, 0)$ respectively.\footnote{All positions are shifted such that site $1$ is at the origin before constructing the FIM in order to fit with the parametrization scheme described in the previous section} Then, for sites $2$ to $N-1$, we generate random coordinates within the sphere of radius $r$ centered on the origin. Once all site positions have been generated, we check that no two sites are closer together than $0.5\units{nm}$ and, if they are, then we generate new site positions for all sites. Although a spacing of $0.5\units{nm}$ would, in practice, likely invalidate the point-dipole approximation and lead to very large nearest neighbour couplings, we choose this as our cutoff so as not to discard too many geometric configurations for a sphere of radius $1\units{nm}$. Furthermore, as can be seen from the vertical dashed lines in Fig.~4 of the main text, the average nearest neighbour coupling for the $r=1\units{nm}$ sphere was around $100\units{meV}$. This tells us that, on average, the nearest neighbour spacings were much closer to $1\units{nm}$, so the precise value of the "too close" cutoff is not important. Once the site positions are acceptable, we then add some disorder to all other base model parameters such that $\eta_i = \eta_{i0} (1+0.1\delta)$ where $\delta$ is a random variable drawn from a normal distribution with a mean of $0$ and a variance of $1$. The values of the various $\eta_0$'s are listed in the Table~\ref{tab:param-vals}.

\section{Validity of Previous Studies}

\begin{figure}
	\includegraphics[scale=0.34]{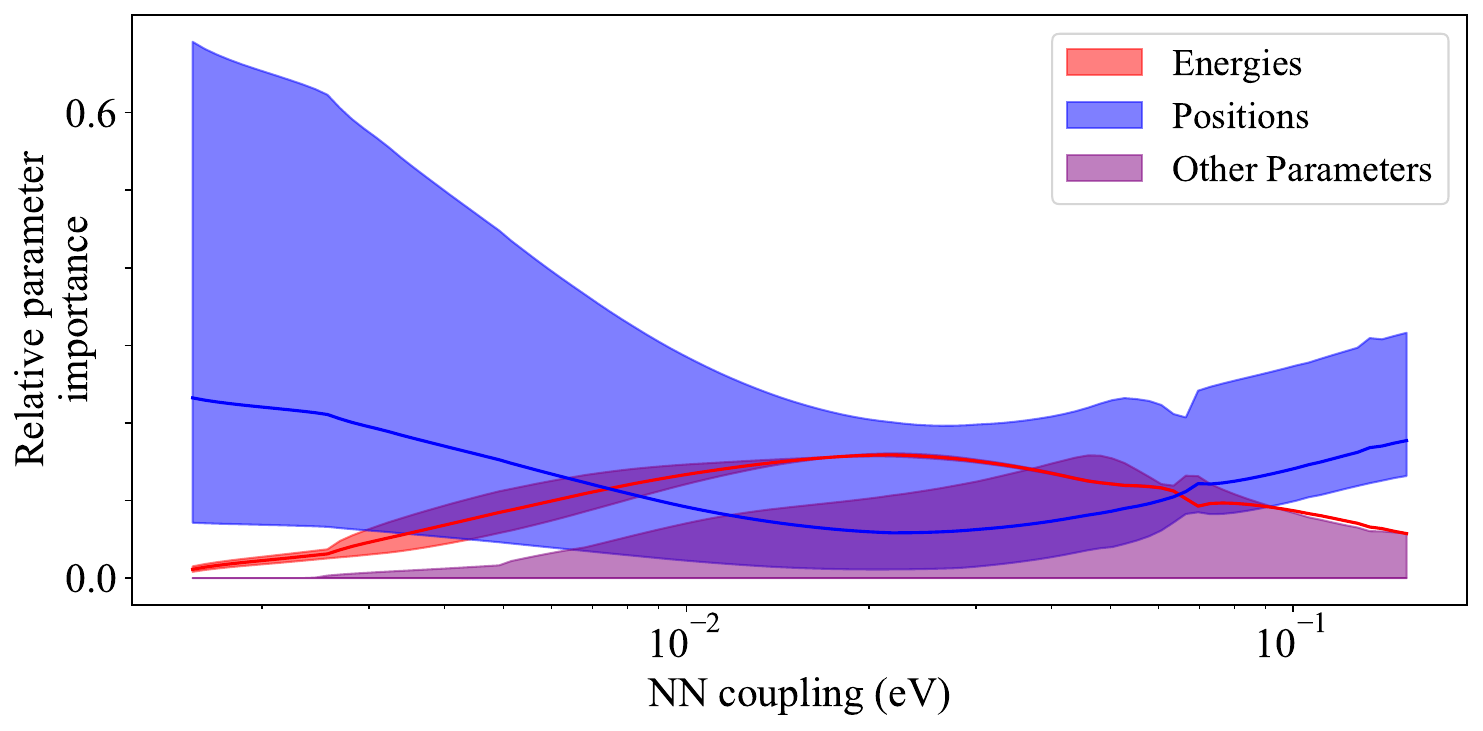}
	\caption{A repeat of the parameter group importance vs nearest neighbour coupling strength plot from the main text, except that the phonon environment has been changed to cause phenomenological site-basis pure dephasing. This leads to clear qualitative differences in the relative importance of the different parameter groups. The `notch' around $0.065\units{eV}$ in the upper bound of the position parameter importance occurs due to a level crossing in the FIM eigenvalues, the details of which are unimportant to the main message of the plot.}
	\label{fig:SI_PureDeph_NoCross}
\end{figure}

Although the main text emphasizes that on-site energies are the important transport parameters in most typical parameter regimes,  we note here that this does not invalidate the previous studies mentioned at the beginning of the main text. Fig.~\ref{fig:SI_PureDeph_NoCross} shows a repeat of the left hand panel of Fig. 4 in the main text with the phonon spectrum changed to an infinite temperature flat spectrum ($J_3$ in the main text). This gives rise to the same phenomenological pure dephasing (in the site basis) considered in many previous studies. With this less realistic form of dephasing, the position parameters have a much stronger influence on the transport properties over a large range of nearest neighbour hopping strengths. Therefore, for many of the pure-dephasing models constructed in previous literature, it is sensible to focus on optimizing the geometric arrangement of sites in order to achieve efficient transport; however, with our more realistic model we find that this is no longer the case.

\section{Dominant Parameter \textit{Combinations}}
\label{SM:param-com}

Strictly speaking, the eigenvectors of the FIM reveal the \textit{linear combinations} of parameters to which the underlying distribution (figure of merit) is most sensitive. In this section we show that there are no clear patterns in the specific FIM eigenvector directions which hold across broad regions of the parameter space. Despite the large variation in these combinations, our methods developed in the main text nonetheless succeed in extracting general, physically informative patterns (e.g.~the dominance of on-site energy components) from these seemingly disparate parameter combinations. 

\begin{figure*}
    \centering
    \includegraphics[scale=0.58]{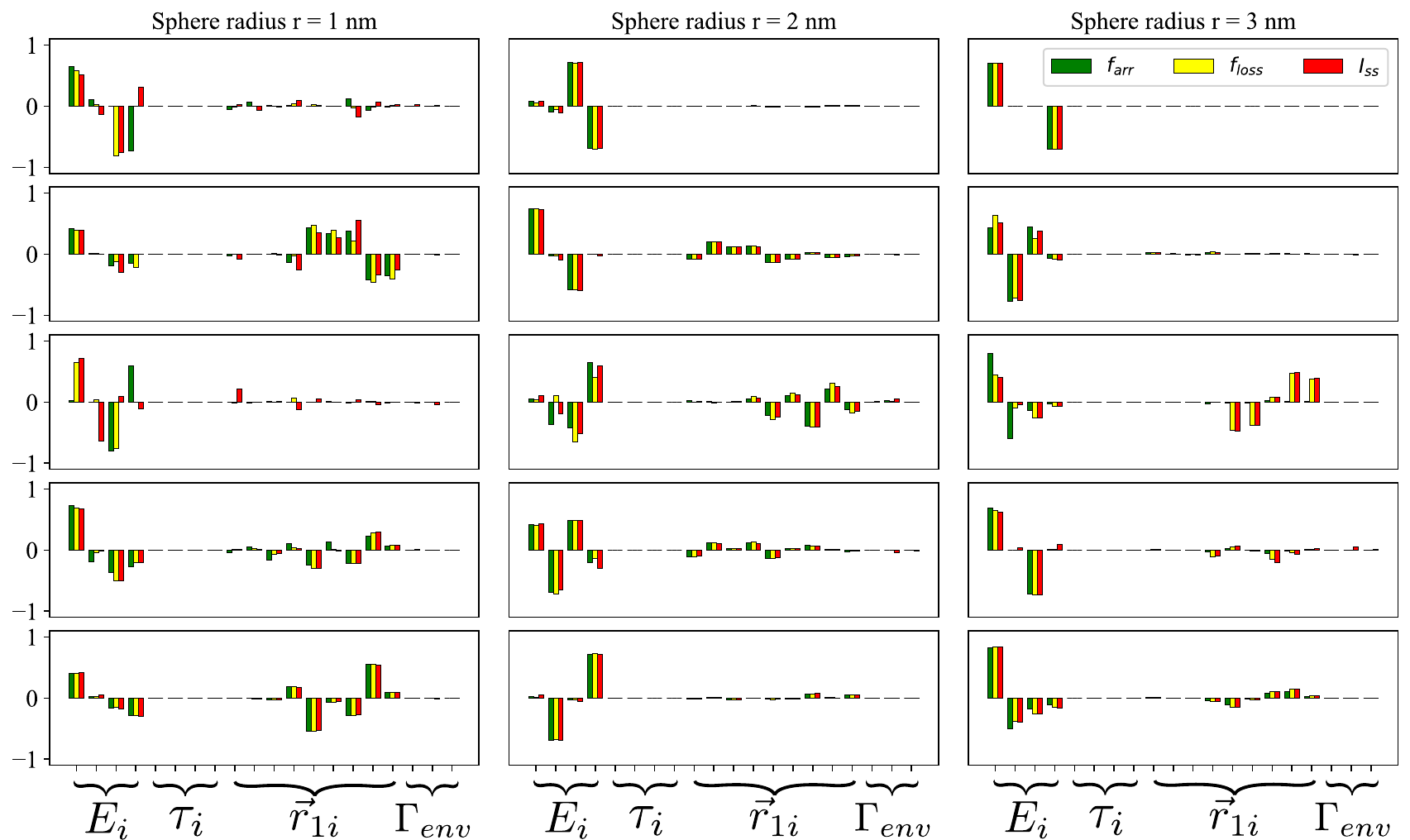}
    \caption{Representative example of the most important parameter combination (largest-eigenvalued eigenvector) from the data used in Fig.~4 of the main text. Five random networks at each sphere radius are shown with all three figures of merit shown for each network. No obvious pattern is visible in the linear parameter combinations and the precise parameter combination which the model is most sensitive to varies significantly for different random networks (i.e. different parameter space regions) and, in some cases, for different figures of merit too. ($\Gamma_{env} = \{\gamma_\text{trap}, \gamma_{ph}, T_{ph}\}$)}
    \label{fig:repr-ex}
\end{figure*}

\begin{figure*}
    \centering
    \includegraphics[scale=0.45]{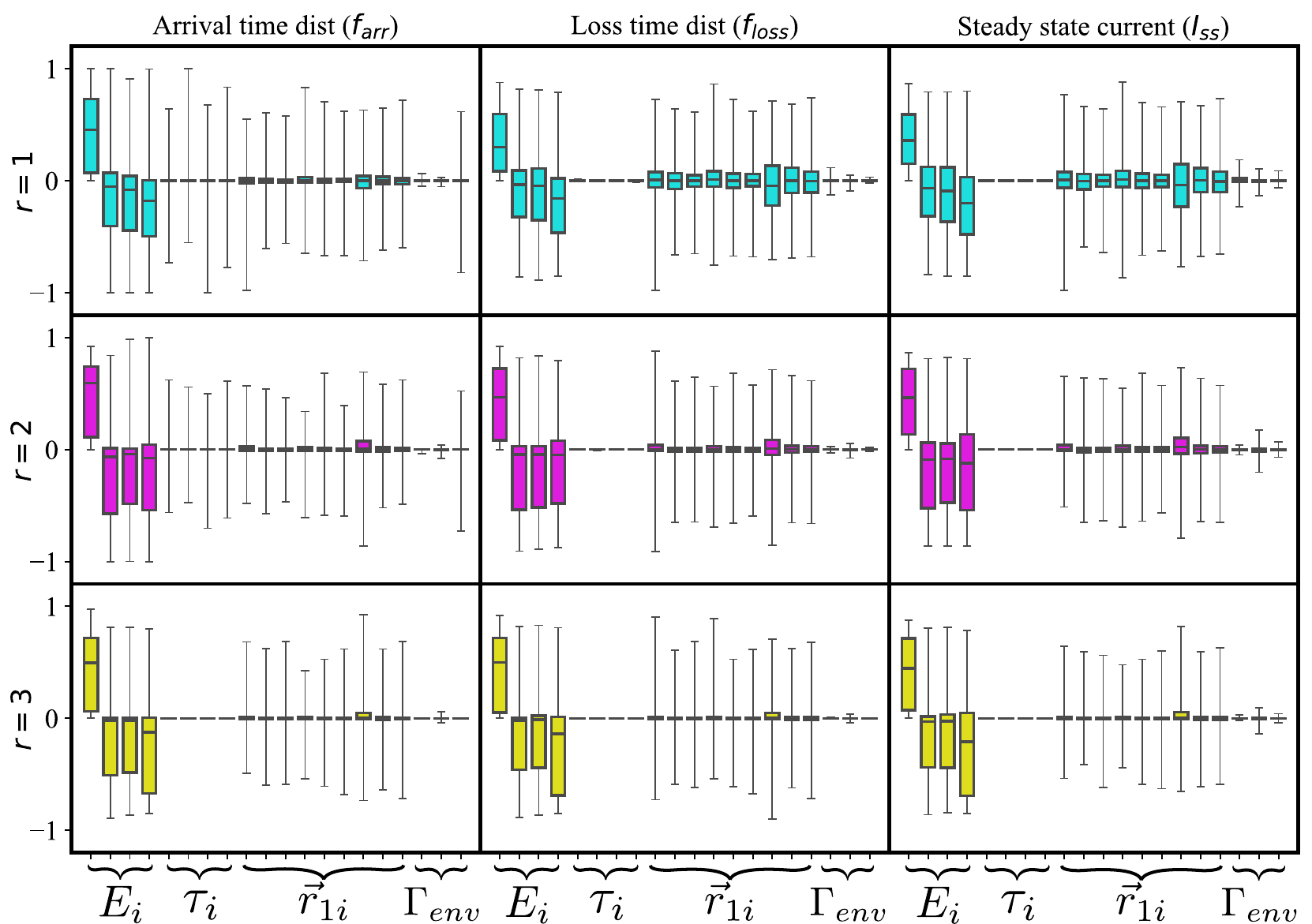}
    \caption{
    Statistical distribution (box plot) of the largest-eigenvalued eigenvector of the FIM (i.e. most important parameter combination) for the randomly generated networks analysed in Fig.~4b of the main text. Rows correspond to different sphere radii (r = 1, 2 \& 3 for top, middle \& bottom rows respectively) whereas columns correspond to the three different figures of merit. Within each box plot, whiskers denote the absolute range of the data while the bottom, middle and upper lines in each box denote the 25th, 50th (i.e. median) and 75th percentiles respectively. Each individual eigenvector is constrained so that the $E1$ component is $>$ 0 for consistency before calculating the statistics (see discussion in SM text). As a result of this constraint, we stress that this figure does \textit{not} imply that downhill energy gradients are always preferable for transport (see Figures~\ref{fig:energy-subspace1}~--~\ref{fig:energy_subspace3} for further discussion).
    }
    \label{fig:box-plot}
\end{figure*}

Before proceeding, it is worth noting that, since any matrix eigenvector can be arbitrarily multiplied by $\pm 1$, the particular direction of each FIM eigenvector component is not informative. Instead it is the relative signs between each component which provides information that can be interpreted in physical terms. For consistency, and without loss of generality, we therefore choose to set the sign of the E1 component as positive for Fig.~\ref{fig:repr-ex}~\&~\ref{fig:box-plot} so that all other parameter components can be seen relative to this constraint. It is important to stress that this does not \textit{per se} imply any preferences for `downhill' energy gradients or other such trends in our results. Furthermore, for the two distribution figures of merit ($f_{arr}$ \& $f_{loss}$) the sensitive parameter combinations are those that `change' the distribution most significantly, therefore it is not strictly possible to speak about parameter combinations which `improve' transport in this context (i.e.~it is possible to change the shape of a distribution without, for example, changing the distribution's expected value significantly).

With these technical details in mind, Fig.~\ref{fig:repr-ex} contains 15 representative examples of the random networks used in Fig.~4b of the main text (5 examples for each sphere radii) and demonstrates that the specific combinations of parameters to which our models are most sensitive (i.e.~the largest-eigenvalued FIM eigenvector) can vary significantly depending on the precise details of the model and individual model parameter values chosen. These important parameter combinations for each random network are clearly distinct and the only apparent pattern across all networks is the dominance of on-site energy and position components -- a finding which was already revealed by the approach developed in the main text. We also intuitively expect that, even for the same model parameters, different figures of merit will sometimes exhibit different key parameter sensitivities -- this is also seen to varying degrees in the random networks in Fig.~\ref{fig:repr-ex}.

For a broader statistical view of this set of important combinations, again with E1 component fixed to be $>$ 0, Fig.~\ref{fig:box-plot} shows a box plot of the components of the most important FIM eigenvector across the 1000 random networks at each sphere radius and for each of the different figures of merit. The `whiskers' on each box show the absolute range of values which that eigenvector component takes across the different transport networks and clearly shows that, due to the broad range of values spanned by most parameter components, the specific important parameter combinations exhibit huge variety across different parameter regimes. Despite this variety, the inter-quartile range (denoted by the span of each rectangular box) clearly demonstrates that the on-site energy components are most widely distributed and therefore dominate the `stiff' parameter subspace of our model's parameter space.

As a final comment of Fig.~\ref{fig:box-plot}, it is notable that the importance of exciton lifetimes is negligible for all loss time distribution and steady state current cases, as well as for the $r=3$ arrival time distribution. This can be explained by our decision to discard any random networks with particularly slow transport performance (steady state time $> 1\units{ns}$) during our random network generating procedure, combined with the fact that the mean lifetime is set at $10\units{ns}$. Therefore, only when the figure of merit is strongly influenced by highly non-equilibrium site populations (i.e.~short time behaviour at small $r$) can the individual lifetimes have a meaningful effect on the figure of merit; in our work, only the arrival time distribution captures this short time behaviour. If the mean lifetime is significantly reduced, lifetime variations will intuitively become more important, though we find that they are still negligible compared with on-site energy and position contributions.

\begin{figure}
    \centering
    \includegraphics[scale=0.42]{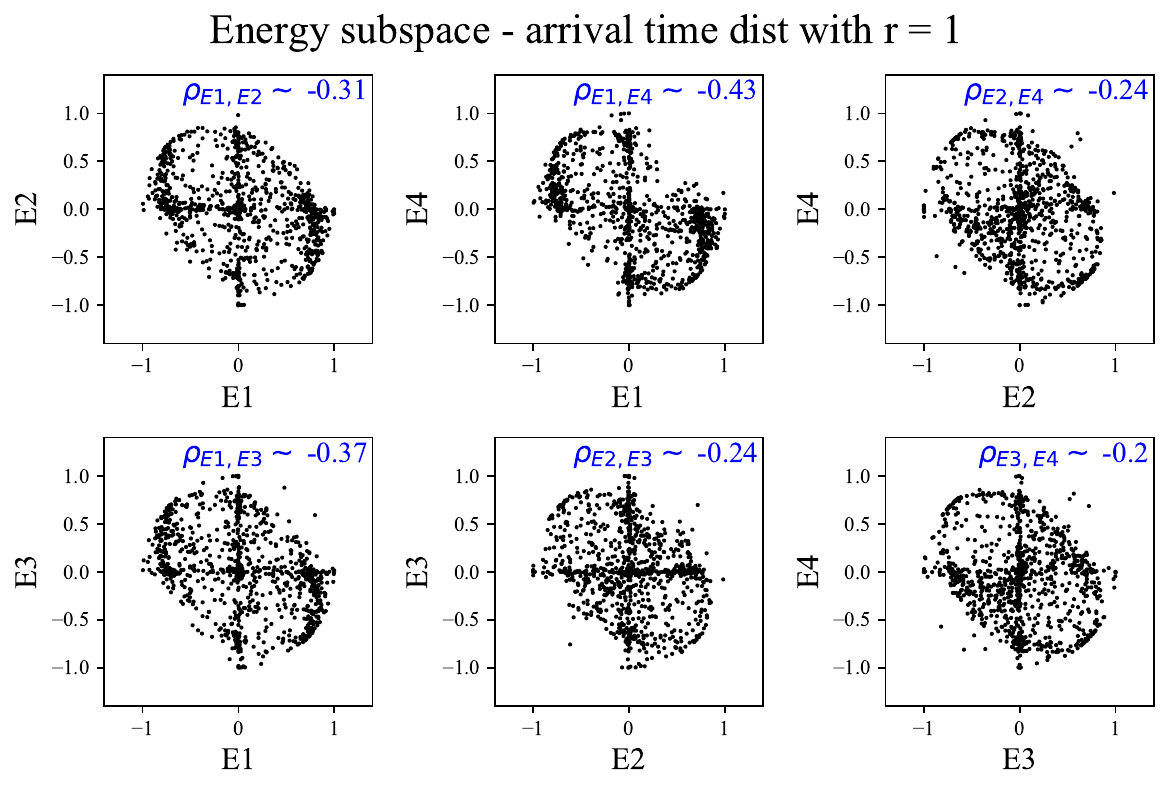}
    \caption{
    Set of 2D slices through the important on-site energy subspace for the $f_{arr}$ distribution and random networks constrained within a sphere of radius $r=1\units{nm}$. Each scatter point represents the contribution to the most important parameter combination in a single random network from the on-site energies on the corresponding $x$ and $y$ axes of that slice. The Pearson correlation coefficient $\rho$ annotated on each slice indicates the weak preference for combinations which lower energies of sites that are closer to the sink site.
    }
    \label{fig:energy-subspace1}
\end{figure}

In order to further demonstrate the variability of important parameter combinations within the on-site energy subspace and their strong dependence on the intricate details of each individual transport network, Fig.~\ref{fig:energy-subspace1} shows the E1--E4 components for each of the most important eigenvectors as a set of 2D slices through this 4D space. From this data we see that there is a slight preference for changes which lower the energies of sites which are located closer to the sink site; however, this is not a particularly strong correlation (as shown by the Pearson correlation coefficient $\rho_{XY}$ annotated on each slice) and there are a significant number of networks in which modifying the energy differences in some other way is more important.

Since the energies in each network are randomly chosen, we can further restrict our correlation calculation to only include the subset networks which already contain an intrinsic energy difference between source and sink sites. Upon doing this, we find $\rho_{E1, E4} \sim -0.37$ (E1~>~E4) and $\rho_{E1, E4} \sim -0.53$ (E1~<~E4) which, by their similarity to the $\rho_{E1, E4}$ value for the full data set (-0.43), demonstrates that the presence of an energy gradient in the network is not a decisive factor in determining the linear combination of on-site energy perturbations to which each network is most sensitive.

\begin{figure}
    \centering
    \includegraphics[scale=0.42]{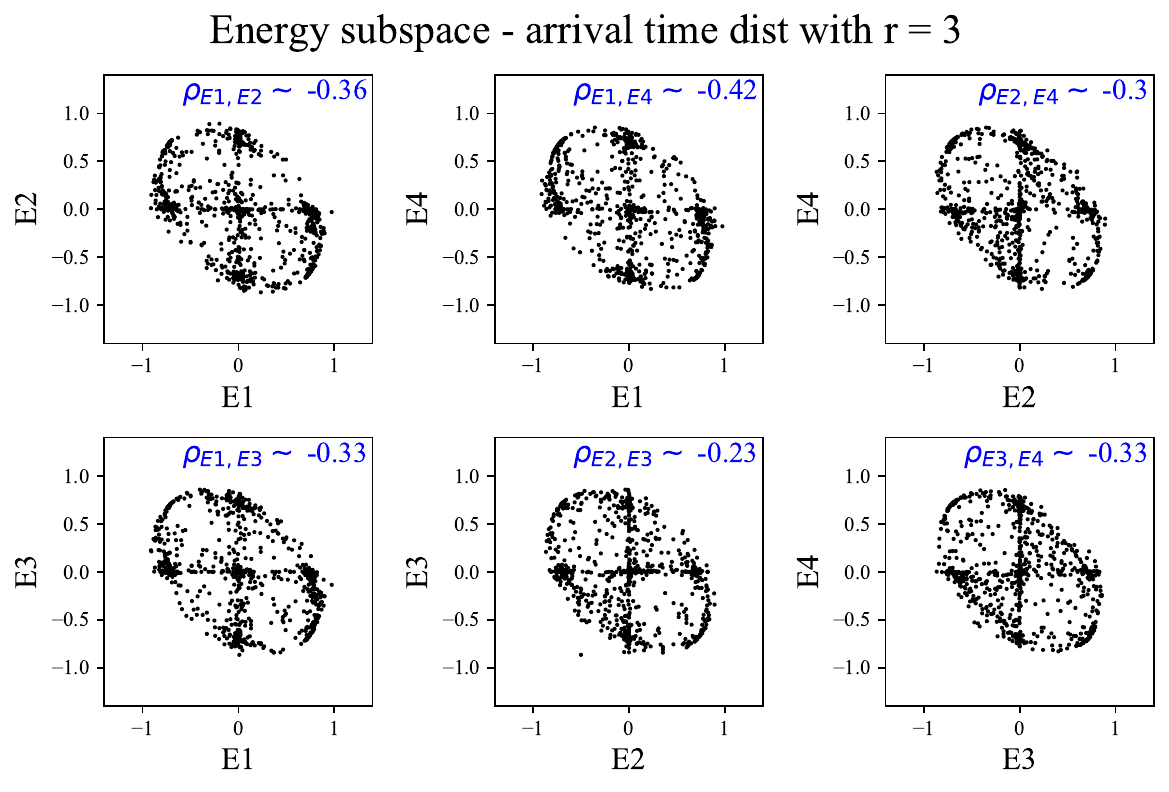}
    \caption{
    Identical plot format to Fig.~\ref{fig:energy-subspace1} for the random networks generated inside a sphere of radius $3\units{nm}$ instead. In this case there is a more prominent clustering of points near the edge of the ellipse, indicative of the increasing importance of on-site energy with sphere radius (i.e. decreasing coherent coupling strength) as found in the main text. 
    }
    \label{fig:energy_subspace2}
\end{figure}

\begin{figure}
    \centering
    \includegraphics[scale=0.42]{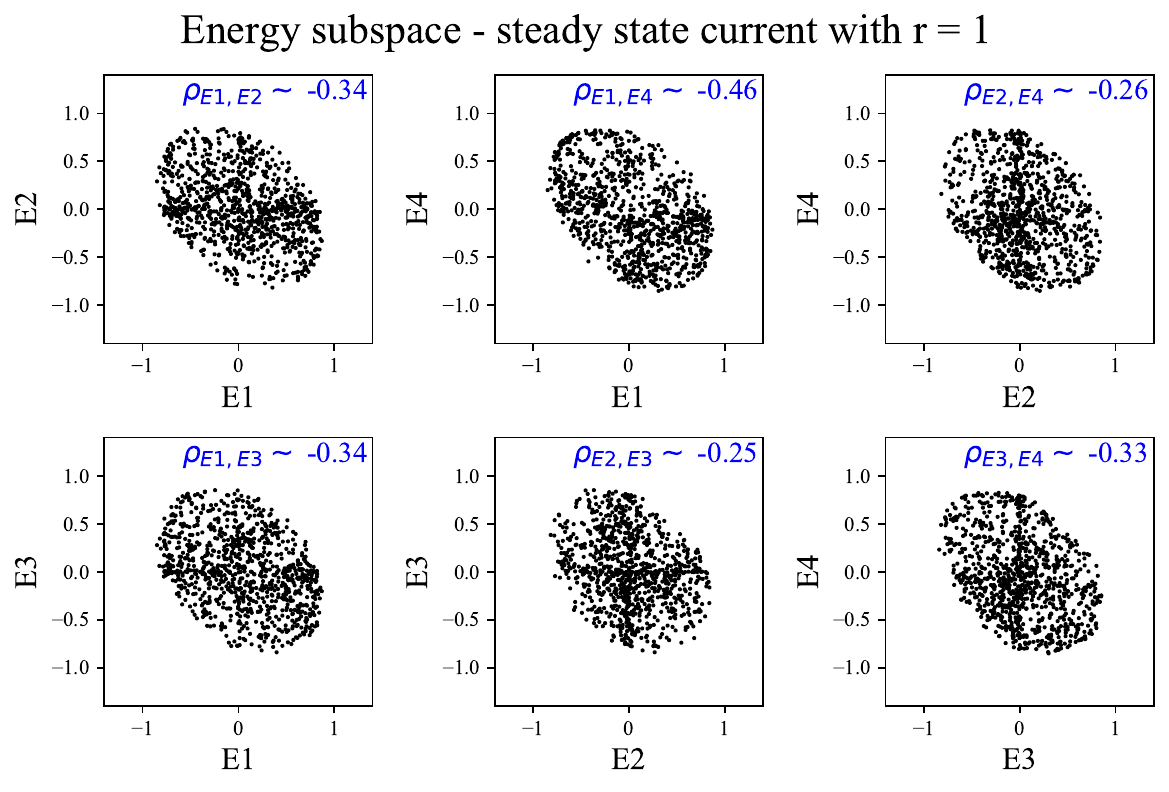}
    \caption{
    Identical plot format to Fig.~\ref{fig:energy-subspace1} for the steady state current figure of merit instead. Here we see a similar weak preference for perturbations which induce energy gradients and that the data points are spread more evenly throughout the space compared with the arrival time distributions used in figures~\ref{fig:energy-subspace1}~\&~\ref{fig:energy_subspace2}.
    }
    \label{fig:energy_subspace3}
\end{figure}

Figures~\ref{fig:energy_subspace2}~\&~\ref{fig:energy_subspace3} show further data for a larger sphere radius and steady state current figure of merit, respectively. It is clear from the similarities between Fig.~\ref{fig:energy-subspace1} and Fig.~\ref{fig:energy_subspace2} that the key parameter combinations are also widely distributed for spheres of larger radius (i.e.~networks with weaker coherent couplings on average). The key difference in the latter figure is the notable clustering of points around the edges of each ellipse, which serves as further confirmation of the increasing on-site energy importance for larger sphere radii (i.e.~weaker coherent coupling), as revealed by the sensitivity analysis techniques presented in the main text.

The alternative steady state current case shown in Fig.~\ref{fig:energy_subspace3} contains a more evenly distributed set of important linear perturbations in the on-site energies, with a similarly weak tendency towards on-site energy gradients. This emphasizes that, even for a simple scalar figure of merit, the important parameter combinations are still highly dependent on the details of each model, without any clear global trends governing the important linear combinations across the entire parameter space.

\section{Varying Number of Sites}

\begin{figure}
	\includegraphics[scale=0.4]{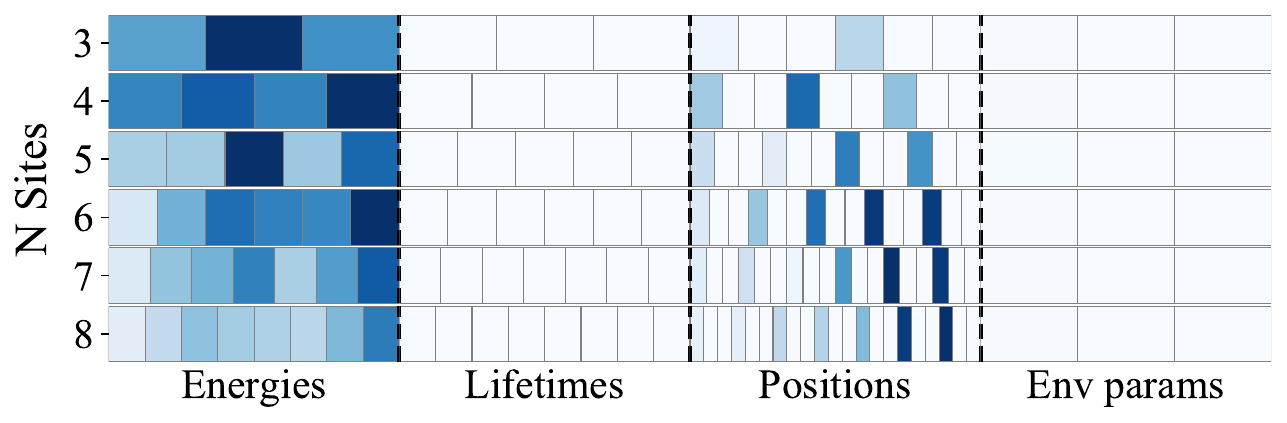}
	\caption{
	Parameter importance for linear chain networks while varying the number of sites in the chain. The source and sink site are kept as the first and last sites in the chain for all $N$. \textit{Nearest neighbour} distances were fixed at $1\units{nm}$ (leading to nearest neighbour couplings around $80\units{meV}$), therefore the distance between source and sink increases with $N$.
	}
	\label{fig:SI_ConstNN_1}
\end{figure}

\begin{figure}
	\includegraphics[scale=0.4]{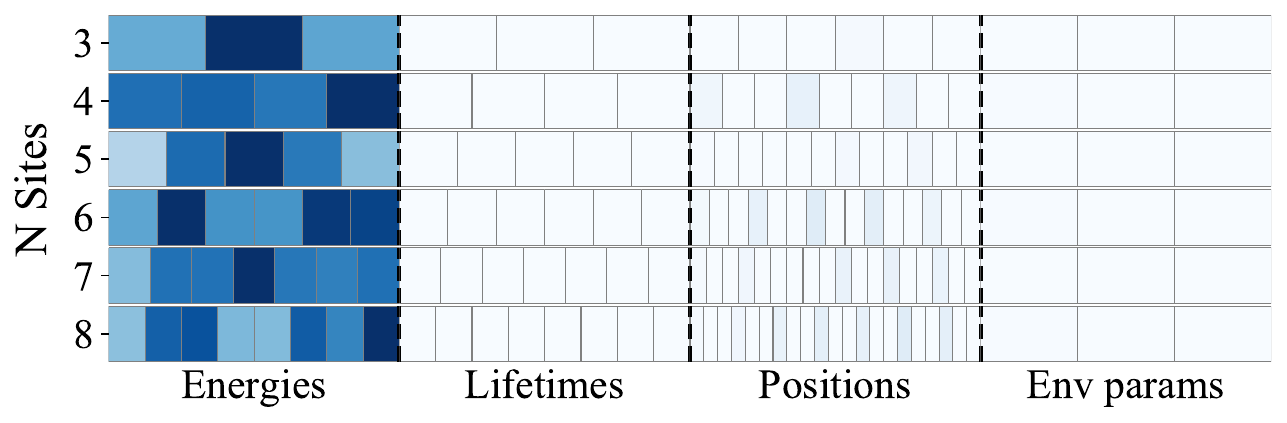}
	\caption{Parameter importance for linear chain networks while varying the number of sites in the chain. All parameters are kept the same as in Fig.~\ref{fig:SI_ConstNN_1} other than the nearest neighbour distances, which were increased to $2\units{nm}$, so that the nearest neighbour coupling is around $10\units{meV}$.}
	\label{fig:SI_ConstNN_2}
\end{figure}

\begin{figure}
	\includegraphics[scale=0.4]{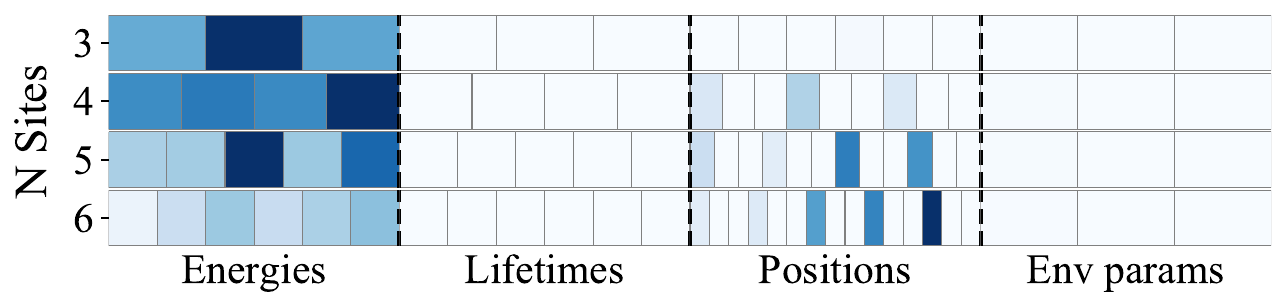}
	\caption{Parameter importance for linear chain networks while varying the number of sites and keeping a constant \textit{source-sink} distance of $5\units{nm}$. The $N-2$ intermediate sites are evenly spaced along the line connecting source to sink, so that higher $N$ leads to stronger nearest neighbour coupling. With six sites and $5\units{nm}$ between source and sink, the sites end up being $<1\units{nm}$ apart which, in practice, would be likely be beyond the limits of the point dipole approximation, therefore we stop at six sites here.}
	\label{fig:SI_ConstSS_1}
\end{figure}

\begin{figure}
	\includegraphics[scale=0.4]{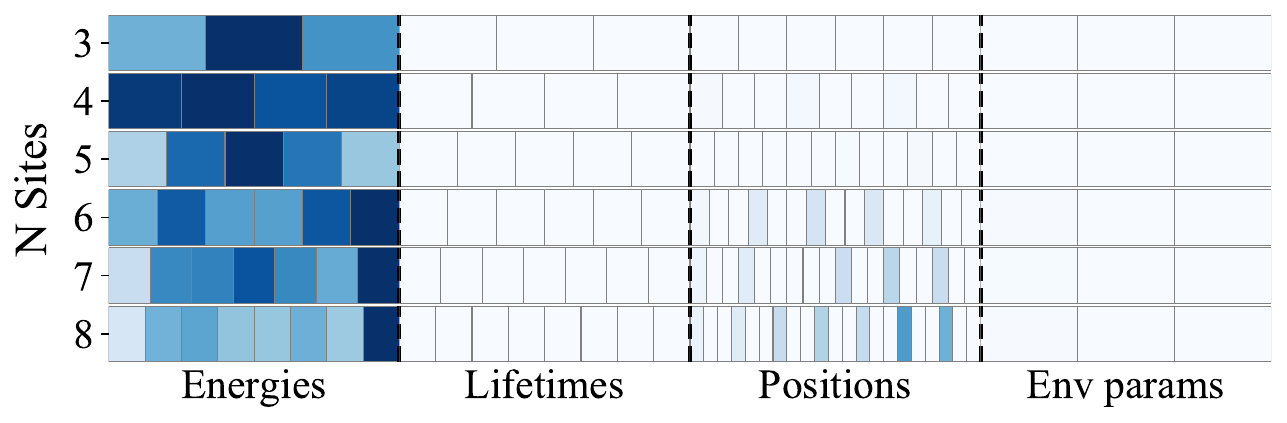}
	\caption{Parameter importance for linear chain while varying the number of sites in the same way as Fig.~\ref{fig:SI_ConstSS_1}. In this case, the source-sink distance is held constant at $10\units{nm}$.}
	\label{fig:SI_ConstSS_2}
\end{figure}

\begin{figure}
	\includegraphics[scale=0.32]{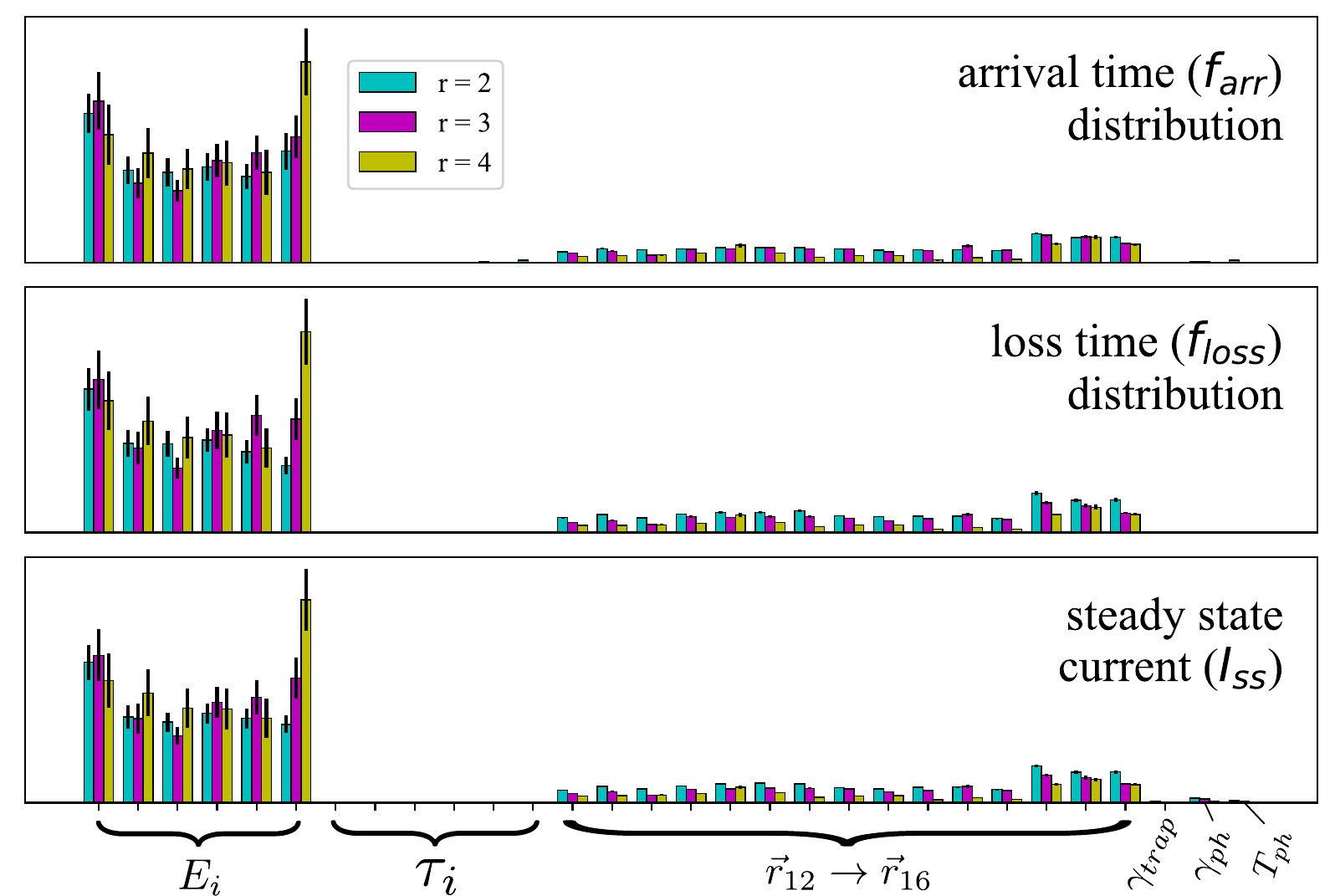}
	\caption{Average parameter importance for 100 randomly generated six-site networks. Generated in the same way as Fig.~4 in the main text.}
	\label{fig:SI_RobustnessN6}
\end{figure}

Figures~\ref{fig:SI_ConstNN_1}-\ref{fig:SI_RobustnessN6} demonstrate that our results are not limited to networks with only four sites. In Figures~\ref{fig:SI_ConstNN_1}~\&~\ref{fig:SI_ConstNN_2} we use a degenerate linear chain network and fix the distance between nearest neighbour sites which leads to an increase in the source-sink distance with increasing number of sites ($N$). Fig.~\ref{fig:SI_ConstNN_1} has nearest neighbour distances fixed at $1\units{nm}$ which leads to strong nearest neighbour couplings ($\sim 80\units{meV}$) and shows that, in all except the three-site network, the positions are somewhat important relative to the site energies. We also see that as $N$ increases, the positions of the end sites in the chain (i.e. those nearest the sink) become much more important. This feature can be explained by a combination of two considerations. Namely, the change in the overlaps of the various system eigenstates with the sites of the chain nearest the sink and the fact that, as the chain length increases, the time evolution lasts longer. This provides more time for the asymmetric phonon-mediated eigenstate transitions to funnel the excitation into the lower energy eigenstates which have a small overlap with the sites near the sink relative to sites nearer the center of the chain. 

For the case shown in  Fig.~\ref{fig:SI_ConstNN_2}, which has a larger nearest neighbour spacing of $2\units{nm}$, leading to weaker nearest neighbour coupling ($\sim 10\units{meV}$), the couplings are weak enough that only the on-site energies are important for all values of $N$ considered here. This agrees with the conclusions of the main text. 

In Figures~\ref{fig:SI_ConstSS_1}~\&~\ref{fig:SI_ConstSS_2}, instead of fixing the nearest neighbour distance, we fix the source-sink distance (i.e. the distance from one end of the chain to the other) and then vary the number of (equally spaced) intermediate sites. This means that as $N$ increases, in contrast to Figures~\ref{fig:SI_ConstNN_1}~\&~\ref{fig:SI_ConstNN_2}, the nearest neighbour coupling strength between sites also increases. Fig.~\ref{fig:SI_ConstSS_1} shows the relative parameter importance using a source-sink separation of $5\,{\rm nm}$. Here we see that as $N$ increases, the resulting increase in nearest neighbour coupling strength causes a transition from on-site energies to site position parameters being the most important for the transport dynamics. For a six-site chain with $5\units{nm}$ source-sink separation, the nearest neighbour couplings are $\sim 0.15\units{eV}$ in our model and so, based on the findings of the main text, it is expected that site positions become important in this parameter regime. This is exactly what we observe in the Figure.

In Fig.~\ref{fig:SI_ConstSS_2}, we now fix the source-sink distance at $10\units{nm}$. In this case, we see that there is a marginal increase in position parameter importance as we increase $N$; however, even for an eight-site chain, the on-site energies are still far more important. This reinforces the conclusion of the left hand panel in Fig.~4 of the main text, since an eight-site chain with $10\units{nm}$ source-sink separation leads to nearest neighbour couplings $\sim 35\units{meV}$. This coupling strength is clearly below the `crossover' region in the main text plot where, on average, position parameters start to become more important that on-site energies.

As well as looking at varying the number of sites in a linear degenerate chain as done above, we can also look at the relative parameter importance for randomly generated networks with more than four sites. Fig.~\ref{fig:SI_RobustnessN6} shows a plot similar to the right hand panel of Fig.~4 in the main text. The difference here is that we instead use six-site transport networks, still with source and sink fixed at opposite poles of a sphere of radius $r$ and all other sites randomly placed within the sphere. All other model parameters also have disorder added to them. As with the main text figure, we restrict ourselves to random networks with relatively short steady state times (< $1\units{ns}$) since we are not interested in cases which exhibit localisation effects or poor transport performance. Furthermore, instead of looking at spheres of radius $\{1, 2, 3\}\units{nm}$, as in the main text, we instead use $r = \{2, 3, 4\}\units{nm}$, so that the six-site networks are not overly compact in their spatial arrangement. Since the relative parameter importance for six-site networks is computationally more expensive to calculate, we only use 100 randomly generated networks here, rather than the $1000$ used in the main text. This figure clearly agrees with our conclusion that the on-site energies are generally more important to the dynamics for the transport model described in the main text. Again, we can see a marginal decrease in the importance of the various position parameters with increasing $r$, but this is negligible in comparison to the importance of the on-site energies.

\section{Scalar Figures of Merit}

Finally, we briefly comment on the differences in the sensitivity analysis procedure for the \textit{scalar} steady state current figure of merit compared with the two distribution figures of merit, $f_{arr}$ and $f_{loss}$, defined in the main text. For a simple `binary choice' scalar figure of merit of the form
\begin{equation}
    f(x, \eta) = P(\eta)^x (1-P(\eta))^{1-x},
\end{equation}
it can be shown that the FIM elements reduce to
\begin{equation}
    g_{\mu\nu} = \bigg( \frac{1}{P} + \frac{1}{1-P} \bigg) \frac{\partial P}{\partial \eta_\mu}  \frac{\partial P}{\partial \eta_\nu},
\end{equation}
and, for this form of FIM, the information geometric picture is no longer valid (since the `embedding space' is now 1D). In this case, the largest FIM eigenvalue is $|\vec{\nabla} P|^2$ and all other eigenvalues are trivially zero due to orthogonality of the eigenvectors. Even though our steady state current figure of merit does not fall into this `binary choice' category, it is still a simple scalar quantity. Therefore, for the parameter importance we simply use the matrix of partial derivatives given by
\begin{equation}
    g_{\mu\nu} = \frac{\partial I_{ss}}{\partial \eta_\mu} \frac{\partial I_{ss}}{\partial \eta_\nu},
\end{equation}
such that the eigenvector of this matrix corresponding to the only non-zero eigenvalue gives us the direction of steepest ascent in parameter space (i.e. $\vec{\nabla} I_{ss}$) and the gradients along all orthogonal directions are zero. The absolute values of the components of this eigenvector in the bare parameter basis then give us the relative importance of each model input parameter for the steady state current. This is a more traditional sensitivity analysis approach as opposed to the information geometry inspired method used for the arrival-time and loss-time distributions described in the main text.


%

\end{document}